\documentclass[fleqn]{annalen}
\usepackage{epsf}
\begin{document}
\newcommand{\volume}{x}              
\newcommand{\xyear}{1999}            
\newcommand{\issue}{x}               
\newcommand{\recdate}{29 July 1999}  
\newcommand{\revdate}{dd.mm.yyyy}    
\newcommand{\revnum}{0}              
\newcommand{\accdate}{dd.mm.yyyy}    
\newcommand{\coeditor}{ue}           
\newcommand{\firstpage}{1}         
\newcommand{\lastpage}{4}          
\setcounter{page}{\firstpage}        
\newcommand{\keywords}{Anderson localization; disordered systems; ergodicity;
	nonlinear systems}
\newcommand{\PACS}{43.40.Ga, 61.43.-j, 63.50.+x, 72.15.Rn}
\newcommand{\shorttitle}{K.A.~Snyder et al., Mode Decay in Excited Nonlinear 
	Systems}
\title{The Influence of Anderson Localization on the Mode Decay of Excited 
	Nonlinear Systems}
\author{K.A.~Snyder$^{1}$ and T.R.~Kirkpatrick$^{2}$} 
\newcommand{\address}{$^{1}$National Institute of Standards and Technology,
	Gaithersburg, MD \,\,20899 \,\, USA\\
  $^{2}$Institute for Physical Science \& Technology and Department of Physics\\
	University of Maryland,
	College Park, MD \,\,20742 \,\,USA}
\newcommand{\email}{\tt kenneth.snyder@nist.gov} 
\maketitle
\begin{abstract}
   A one-dimensional system of masses with nearest-neighbor 
interactions and periodic boundary conditions 
is used to study mode decay and ergodicity in nonlinear, 
disordered systems.  
The system is given an initial periodic displacement, and 
the total system energy within a specific frequency channel is measured as a 
function of time.  
Results indicate that the rate of mode decay at early times increases 
when impurities are added.  However, for 
long times the rate of mode decay decreases 
with increasing impurity mass and impurity concentration.  
This behavior at long times can be explained by Anderson localization effects
and the nonergodic response of the system.
\end{abstract}

\section{Introduction}
The transition from quasi-periodic to ergodic behavior in nonlinear systems 
has been an active field of research since the seminal work of Fermi, 
Pasta, and Ulam (FPU)~\cite{FPU}.  The behavior of the FPU system 
has been discussed in 
great detail~\cite{Ford92,Casetti97}, and given the degree of 
nonlinearity and initial energy density, one can estimate the time required 
before the system becomes ergodic.   However, little is known about the effects 
that impurities have upon the transition to ergodic behavior in nonlinear 
systems. 
Since mode coupling in nonlinear systems occurs through interactions, one 
expects that the presence of 
impurities will hasten the transition to ergodic behavior.
This expectation is, however, in general not correct due to Anderson 
localization effects~\cite{Anderson}.

Reported here are the results from a numerical experiment using 
a one-dimensional system of masses with nonlinear nearest neighbor forces 
and periodic boundary conditions.  The 
displacement of each mass is sampled over a finite interval of time and 
the energy within all frequencies is calculated for each mass.  The 
total energy in a single mode is conserved in 
harmonic systems both with and without 
impurities.  Therefore, for the nonlinear 
systems, the effects of the impurities upon the mode decay can be 
compared directly to the systems without impurities.

\section{Numerical Experiment}
The system used in this experiment, which corresponds to the FPU $\beta$-model,
is composed of $N$ masses with periodic boundary conditions, and  
with unit equilibrium spacing. 
The masses undergo a displacement $u(t)$, and the 
$i$-th mass $m_i$ interacts through nearest neighbor forces:
\begin{equation}
   m_i \ddot{u}_i = -\kappa
	\left[\left(u_{i+1}-u_{i}\right)-\left(u_{i}-u_{i-1}\right)\right]-
	\beta
	\left[\left(u_{i+1}-u_{i}\right)^3-\left(u_{i}-u_{i-1}\right)^3\right]
  \label{force_eqn}
\end{equation}
For this experiment, the coupling constants $\kappa=\beta=1$.
A pure system is composed of masses $m^o=1$, and disorder is achieved by 
randomly changing a number $N_I$ of the masses to a second value.
The time integration is performed using a 4-th order 
predictor-corrector algorithm.  Further details will 
be given elsewhere.

The initial condition is a unit amplitude, zero-velocity sinusoidal 
displacement with wavelength $\lambda=32$.  Time is scaled by the harmonic 
frequency $\omega^o=2\pi/\lambda$.
At certain intervals, the time-dependent displacement of each mass is 
transformed to frequency space, and these data $u_k(x_i)$ 
are used to calculate 
the modal energy $E_\omega=E(\omega^o)$ and mass energy 
$E_i=E(x_i)$ by summing over masses and frequencies, respectively.

\section{Results}
The systems consist of $N=256$ masses, which is a 
size consistent with systems used elsewhere~\cite{Casetti97}.  
Time is expressed as the dimensionless quantity $\omega^o t/2\pi$ 
which is equivalent to the number of harmonic cycles, 
and the modal energy is scaled to $E(\omega^o t/2\pi=32)$.  The 
error bars shown in the figures represent the estimated standard deviation 
in the mean.

\begin{figure}[htb]
   \includegraphics{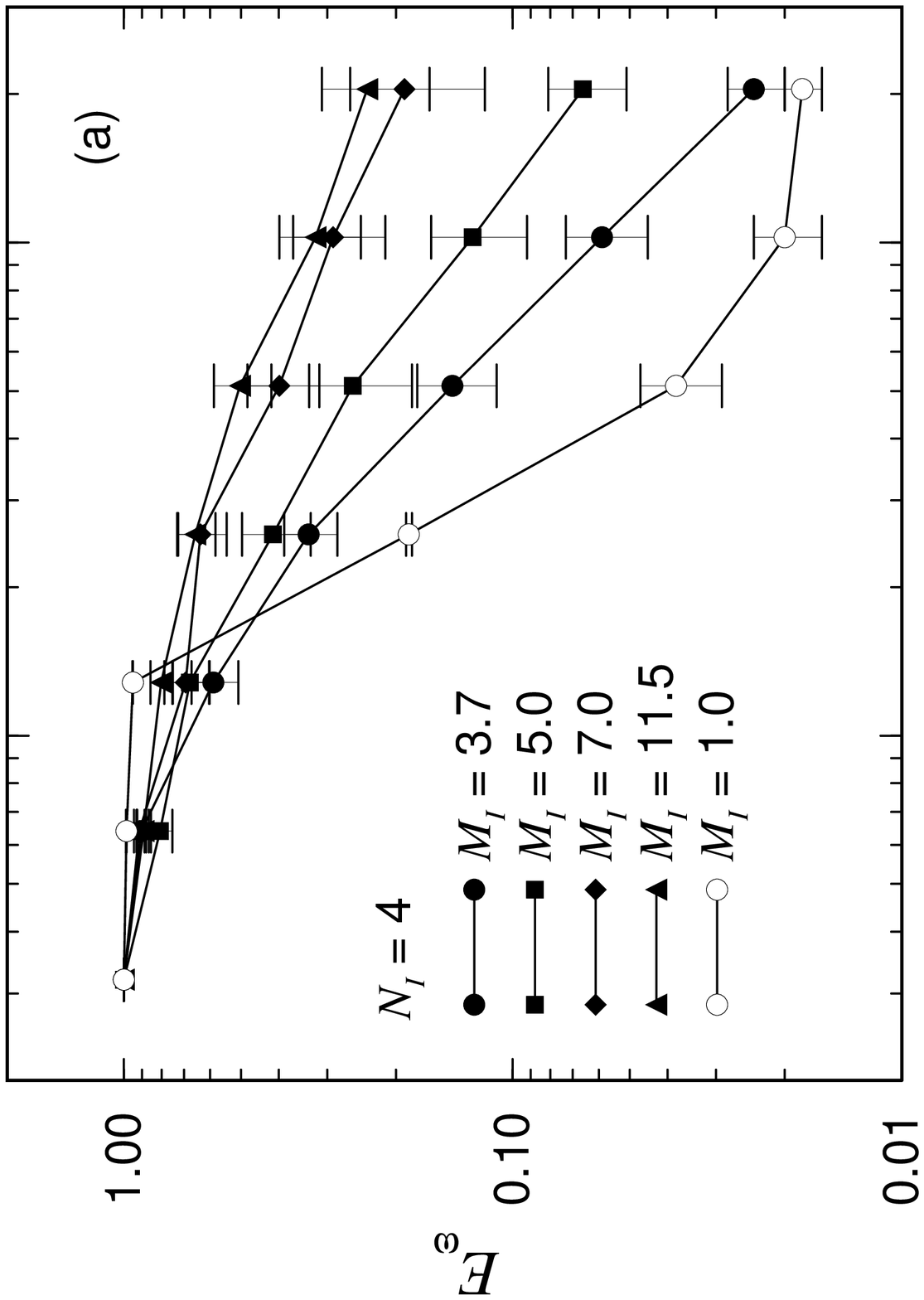}
   \includegraphics{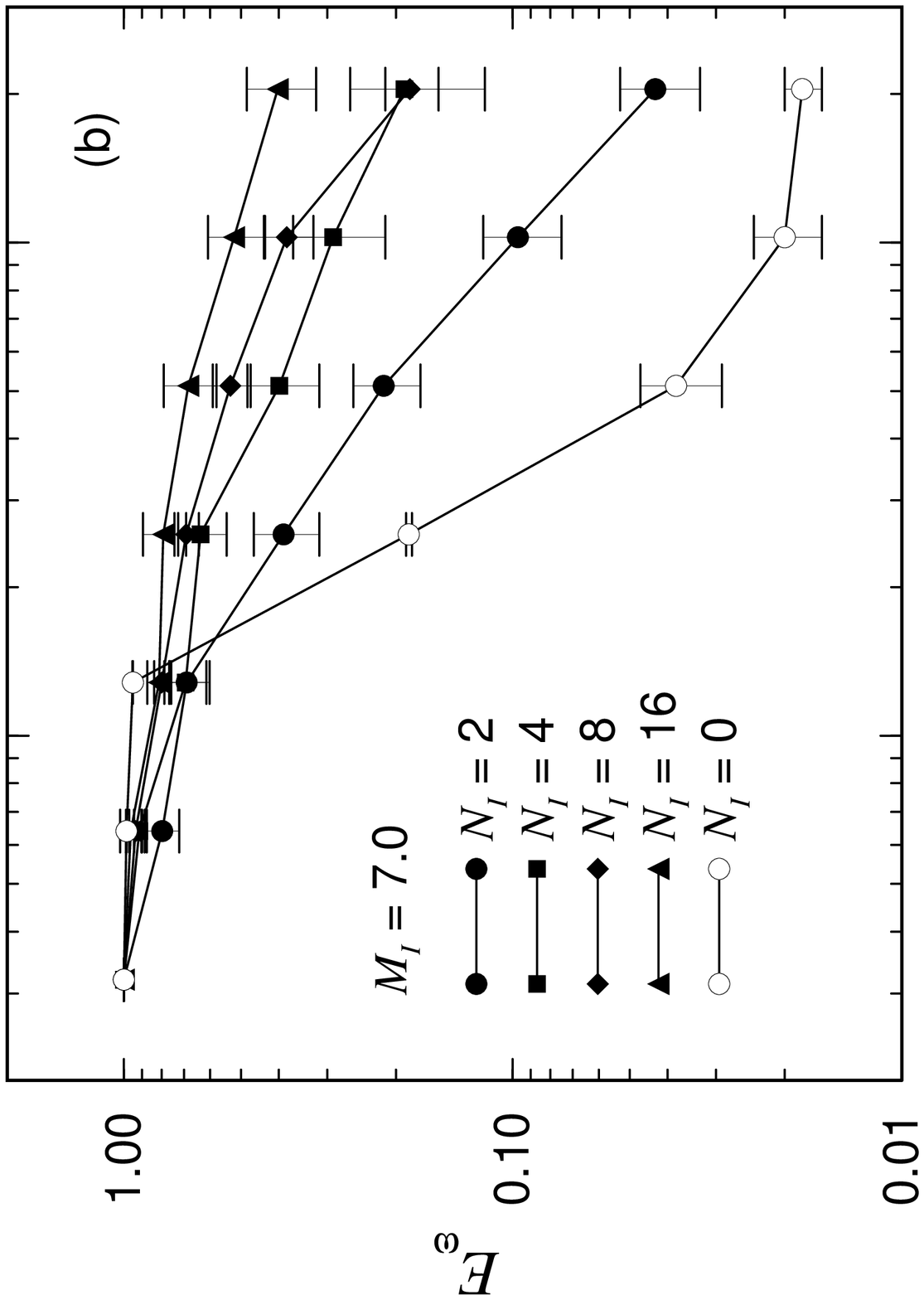}
   \includegraphics{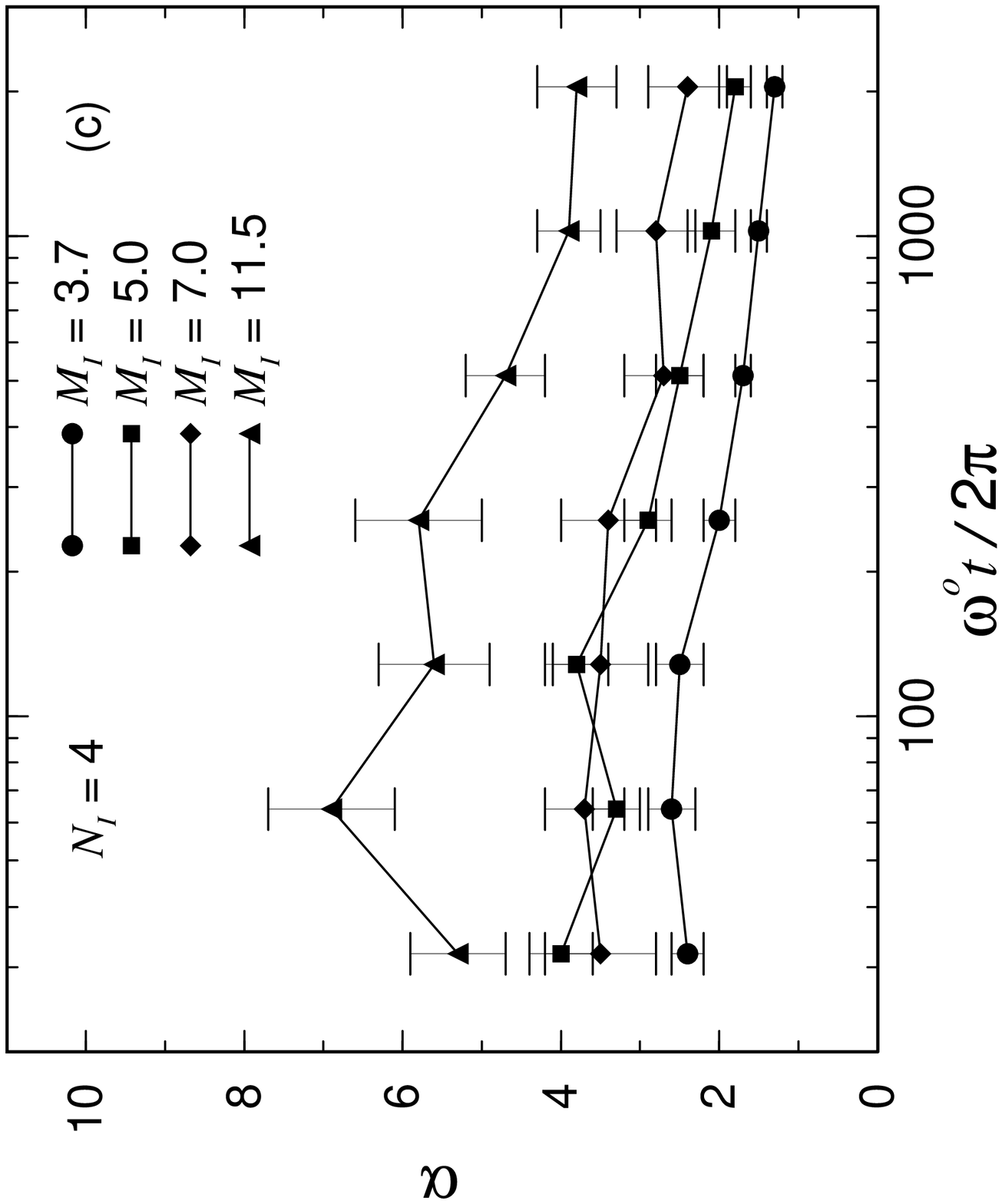}
   \includegraphics{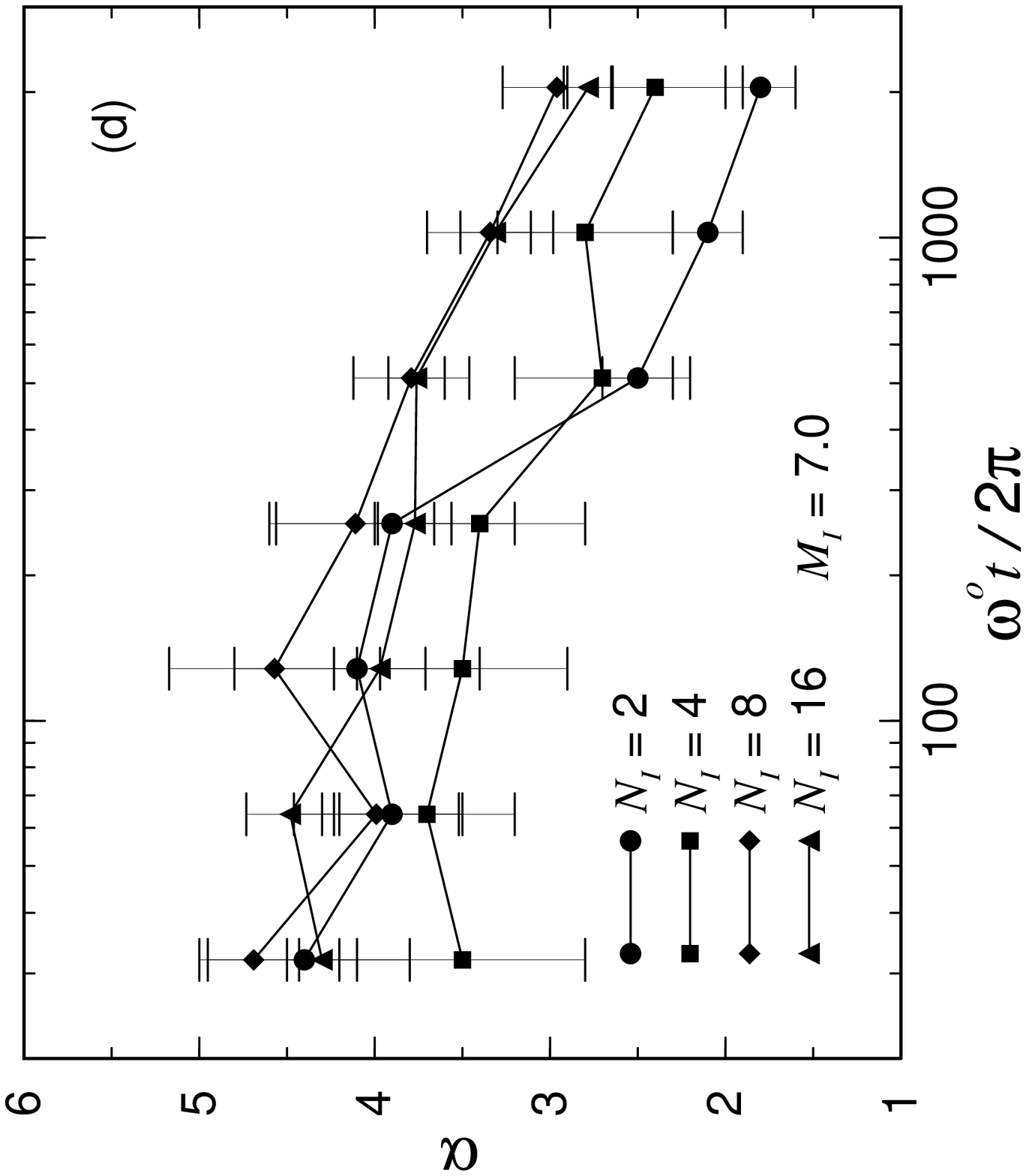}
  \vspace{4.0in}
  \caption{Effect of impurity mass $M_I$ and number of impurities $N_I$  on the 
	mode energy $E_\omega$ and relative energy $\alpha$ at an impurity. 
	In (a) and (c) the number of impurities $N_I$ is 4.  In (b) and (d) the 
	impurity mass $M_I$ is 7.  The open circles are for the nonlinear 
	system without impurities.  Error bars represent the estimated standard 
	deviation of the mean.} 
  \label{HeavyMode_fig}
\end{figure}

The initial systems investigated contained impurity masses 
$M_I > m^o$.   Comparisons of the effects upon 
modal decay as a function of either impurity mass $M_I$ or 
number of impurities $N_I$
are shown in Figs.~\ref{HeavyMode_fig} (a) and (b); 
the pure nonlinear system is denoted by open circles.
As expected, the addition of impurities increases the initial rate 
at which the harmonic mode $\omega^o$ 
decays.  However, for all the combinations of 
impurity mass and concentration shown, the 
long time decay is {\it slower} than for the system without impurities.  
Further, increases in impurity mass and impurity concentration retard
the long time decay.

The long time behavior of the systems containing heavy impurities can 
be explained by the nonergodic behavior within the system.  Systems 
containing impurities undergo an Anderson transition.  The 
resulting excited modes are localized,
concentrating energy near the impurities.  Since the impurities are 
heavier than $m^o$, their oscillation amplitudes 
are smaller, resulting in a smaller nonlinear contribution to the 
energy at the impurity.  

As a demonstration of the localization of energy at the impurities, 
the mass energy $E_i$ is calculated at each impurity.  The average 
energy at an impurity $\langle E_i\rangle_I$ is compared to the 
average energy at all the masses $\langle E_i\rangle_N$.
The ratio $\alpha=\langle E_i\rangle_I/\langle E_i\rangle_N$, which is 
interpreted as a measure of the extent to which ergodicity has been achieved, 
is shown in Figs.~\ref{HeavyMode_fig}(c) and (d).

Figures~\ref{HeavyMode_fig}(c) and (d) show that the relative energy at an 
impurity increases with increasing impurity mass, and 
is somewhat insensitive to changes in the impurity concentration, 
respectively.  
Therefore, the total energy located at impurities 
is proportional to the mass and the number of impurities.  
This explains why the 
modal decay rate decreases as either the impurity mass or the impurity 
concentration increases.

\begin{figure}[htb]
   \includegraphics{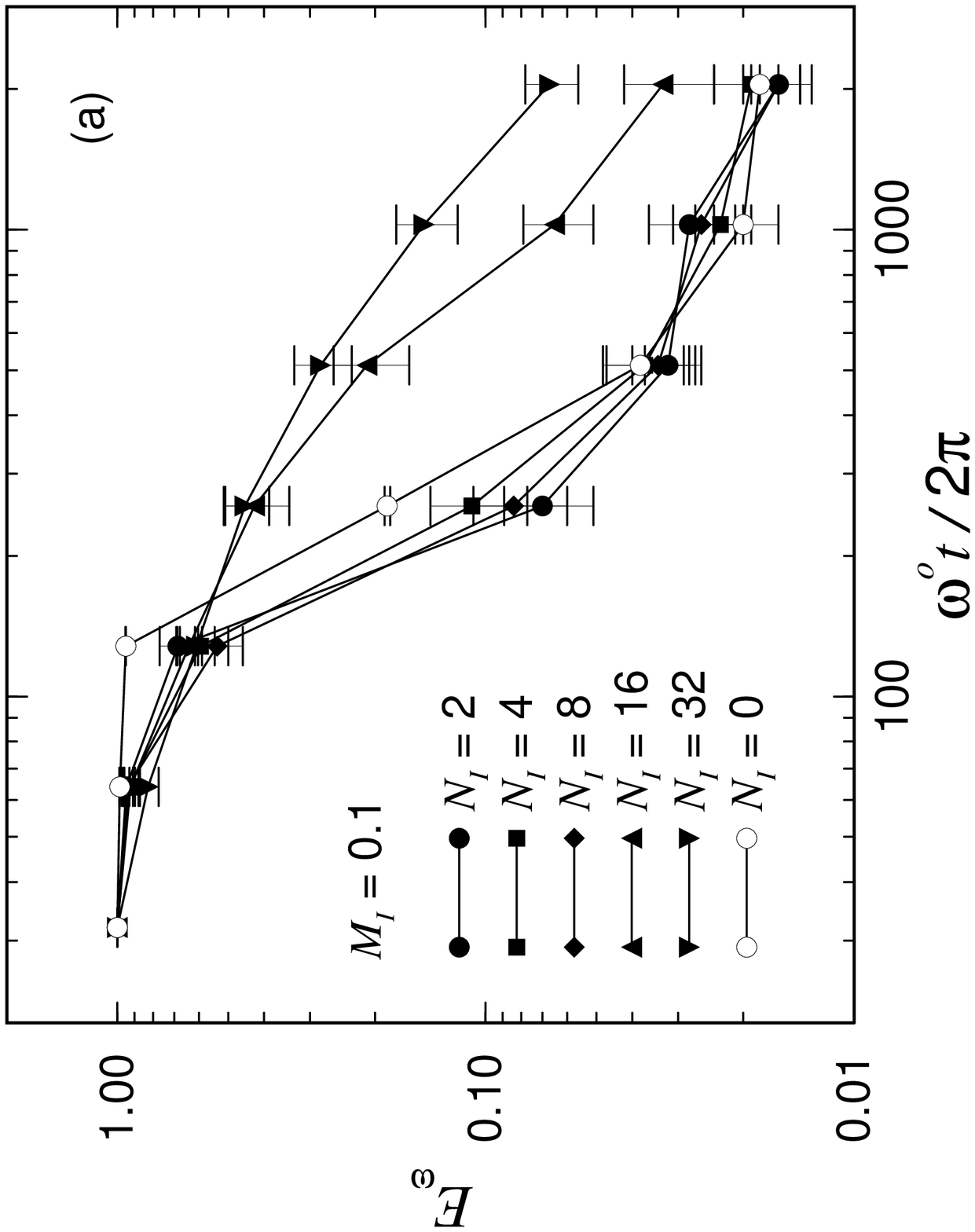}
   \includegraphics{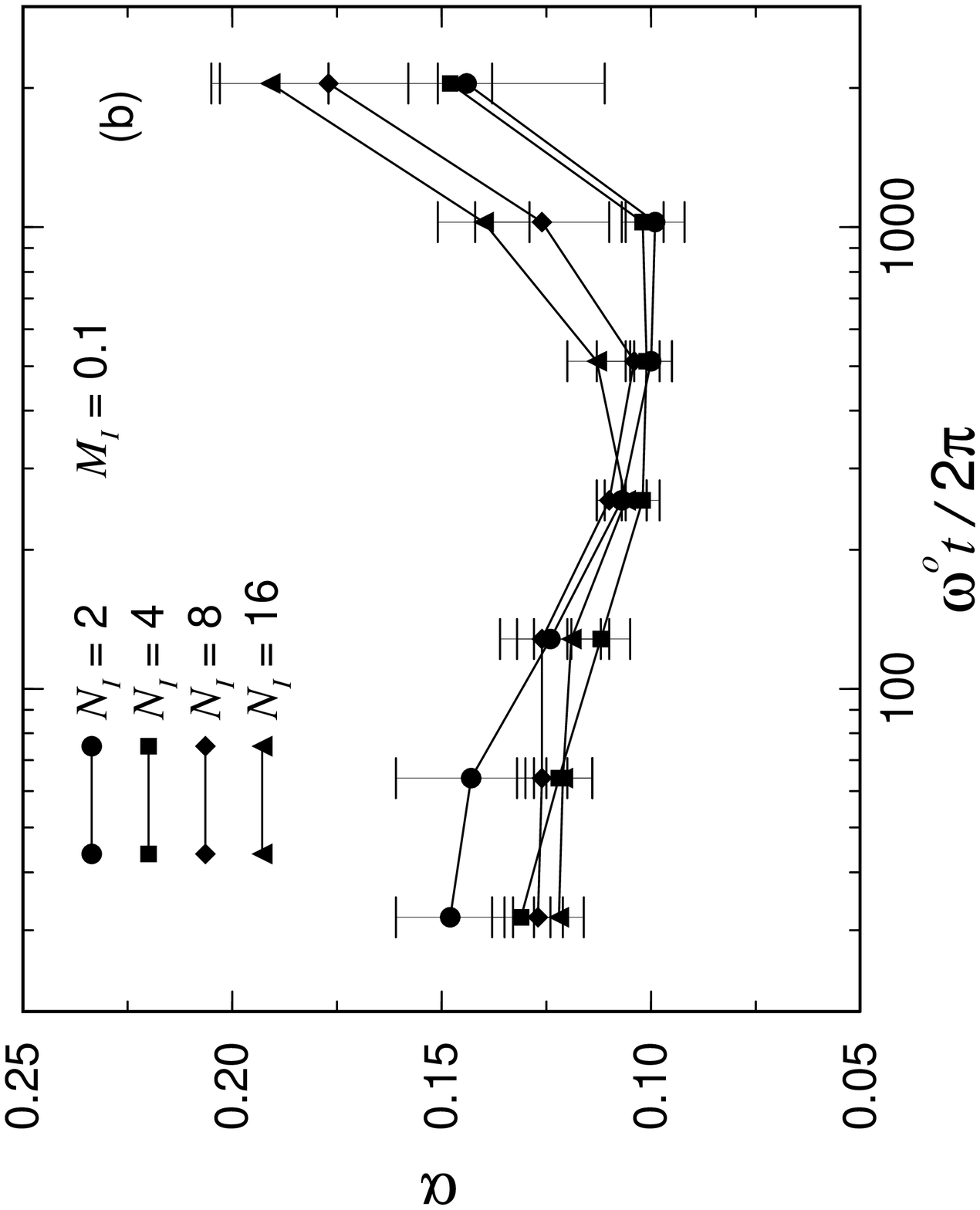}
  \vspace{2.25in}
  \caption{The modal energy $E_\omega$ and the relative energy at the 
	impurities $\alpha$ for impurities of mass $M_I=0.1$.}
  \label{Light_fig}
\end{figure}

Based upon the arguments given above, impurities with masses that are 
lighter than $m^o$ should have a different effect upon mode 
decay.  Since the impurities are lighter, the localized energy will 
create large oscillations, which should enhance nonlinear interactions and, 
hence, enhance the rate of mode decay.  
The results of that experiment for impurity mass $M_I=0.1$ 
are shown in Fig.~\ref{Light_fig}.
For low concentrations, the mode decay shown in Fig.~\ref{Light_fig}(a) 
is faster than for the pure 
system, as expected.  However, at sufficiently high concentration, the 
mode decays slower than the pure system when the 
average spacing between impurities is 
one half the initial wavelength.

The relative energy $\alpha$ located at a light impurity shown in 
Fig.~\ref{Light_fig}(b) is 
less than unity.  The light impurities apparently expel energy due to the 
large oscillation amplitudes, hastening mode decay.  Above a critical 
concentration, it is conjectured that the light impurities interact in such a 
way as to prolong the mode decay.

\section{Conclusion}
For the simulation times studied here, the long time mode decay in 
disordered nonlinear systems seems to be 
controlled by disorder and Anderson localization effects.  
For either heavy or 
light impurities the energy becomes localized, and the assumption 
of ergodic behavior is not valid.  
Further, the response of the system 
depends upon whether the impurities are heavier or lighter than the 
pure system.  
Heavier impurities consume energy and, because of the 
smaller oscillation amplitudes, release their energy through nonlinear 
interactions very slowly.  
Lighter impurities appear to expel energy through 
large oscillation amplitudes, hastening nonlinear interactions and 
mode decay at low concentrations, but behave in a manner similar to 
heavy impurities at high concentrations, possibly due to Anderson 
localization effects.

\vspace*{0.25cm} \baselineskip=10pt{\small \noindent The authors 
   would like to thank Jack Douglas of the Polymers Division (NIST) 
   for his insightful comments and discussion.  The authors would also like to 
   acknowledge the support of the National Science Foundation through 
   Grant DMR~99-75259.}
%
%
%
%
%
%
%
%
%
%
%
%

\end{document}